\documentclass[prl,twocolumn,superscriptaddress]{revtex4}

\usepackage{graphicx}

\usepackage[normalem]{ulem} 
\usepackage{color} 

\begin{document}

\title{Rare earth spin ensemble magnetically coupled to a superconducting resonator}

\author{P.~Bushev}
\affiliation{Physikalisches Institut, Karlsruhe Institute of Technology, D-76128 Karlsruhe, Germany}

\author{A.~K.~Feofanov}
\affiliation{Physikalisches Institut, Karlsruhe Institute of Technology, D-76128 Karlsruhe, Germany}

\author{H.~Rotzinger}
\affiliation{Physikalisches Institut, Karlsruhe Institute of Technology, D-76128 Karlsruhe, Germany}

\author{I.~Protopopov}
\affiliation{Institut f\"{u}r Nanotechnologie, Karlsruhe Institute of Technology, D-76021 Karlsruhe, Germany}

\author{J.~H.~Cole}
\affiliation{Institut f\"{u}r Theoretische Festk\"{o}rperphysik, Karlsruhe Institute of Technology, D-76128 Karlsruhe, Germany}
\affiliation{Applied Physics, School of Applied Sciences, RMIT University, Melbourne 3001, Australia}

\author{C.~M.~Wilson}
\affiliation{Microtechnology and Nanoscience, MC2, Chalmers University of Technology, SE-412 96 G\"{o}teborg, Sweden}

\author{G.~Fischer}
\affiliation{Physikalisches Institut, Karlsruhe Institute of Technology, D-76128 Karlsruhe, Germany}

\author{A.~Lukashenko}
\affiliation{Physikalisches Institut, Karlsruhe Institute of Technology, D-76128 Karlsruhe, Germany}

\author{A.~V.~Ustinov}
\affiliation{Physikalisches Institut, Karlsruhe Institute of Technology, D-76128 Karlsruhe, Germany}
\affiliation{DFG-Center for Functional Nanostructures (CFN), D-76128 Karlsruhe, Germany}

\date{\today}

\begin{abstract}
Interfacing superconducting quantum processors, working in the GHz frequency range, with optical quantum networks and atomic qubits is a challenging task for the implementation of distributed quantum information processing as well as for quantum communication. Using spin ensembles of rare earth ions provides an excellent opportunity to bridge microwave and optical domains at the quantum level. In this letter, we demonstrate magnetic coupling of Er$^{3+}$ spins doped in a Y$_{2}$SiO$_{5}$ crystal to a high-Q coplanar superconducting resonator.
\end{abstract}

\pacs{42.50.Fx, 76.30.Kg, 03.67.Hk, 03.67.Lx, 76.30.-v}

\keywords{Cooperative phenomena in quantum optical systems, Superconducting qubits, Quantum communication, Quantum computation architectures and implementations, EPR in condensed matter}

\maketitle


Quantum communication is a rapidly developing field of science and technology which allows the transmission of information in an intrinsically secure way~\cite{Gisin2007}. As well as its classical counterpart, a quantum communication network can combine various types of systems which transmit, receive, and process information using quantum algorithms~\cite{Kimble2008}. For example, the nodes of such network can be implemented by superconducting (SC) quantum circuits operated in the GHz frequency range~\cite{Clarke2008}, whereas fiber optics operated at near-infrared can be used to link them over long distances. For reversible transfer of quantum states between systems operating at GHz and optical frequency ranges one must use a hybrid system~\cite{Molmer2008}. Spin ensembles coupled to microwave resonator represents one of the possible implementations of such a system~\cite{Verdu2009,Imamoglu2009}. The collective coupling strength of a spin ensemble is increased with respect to a single spin by the square root of the number of spins. Transparent crystals doped with paramagnetic ions often possess long coherence times~\cite{Sun2006,Bertaina2007,Awschalom2008}, and the collective coupling has been recently demonstrated with NV-centers in diamond~\cite{Bertet2010,Shuster2010,Amsuss2011} and (Cr$^{3+}$) ions in ruby~\cite{Shuster2010}.

In this letter, we report on the measurement of a spin ensemble of Erbium ions in a crystal, magnetically coupled to a high-Q coplanar SC resonator. The Er$^{3+}$ ions are distinct from other spin ensembles due to their optical transition at telecom C-band, i.e. inside the so-called ``Erbium window'' at 1.54 $\mu$m wavelength, and long measured optical coherence time.

\begin{figure}[htb]
    	\includegraphics[width=0.9\columnwidth]{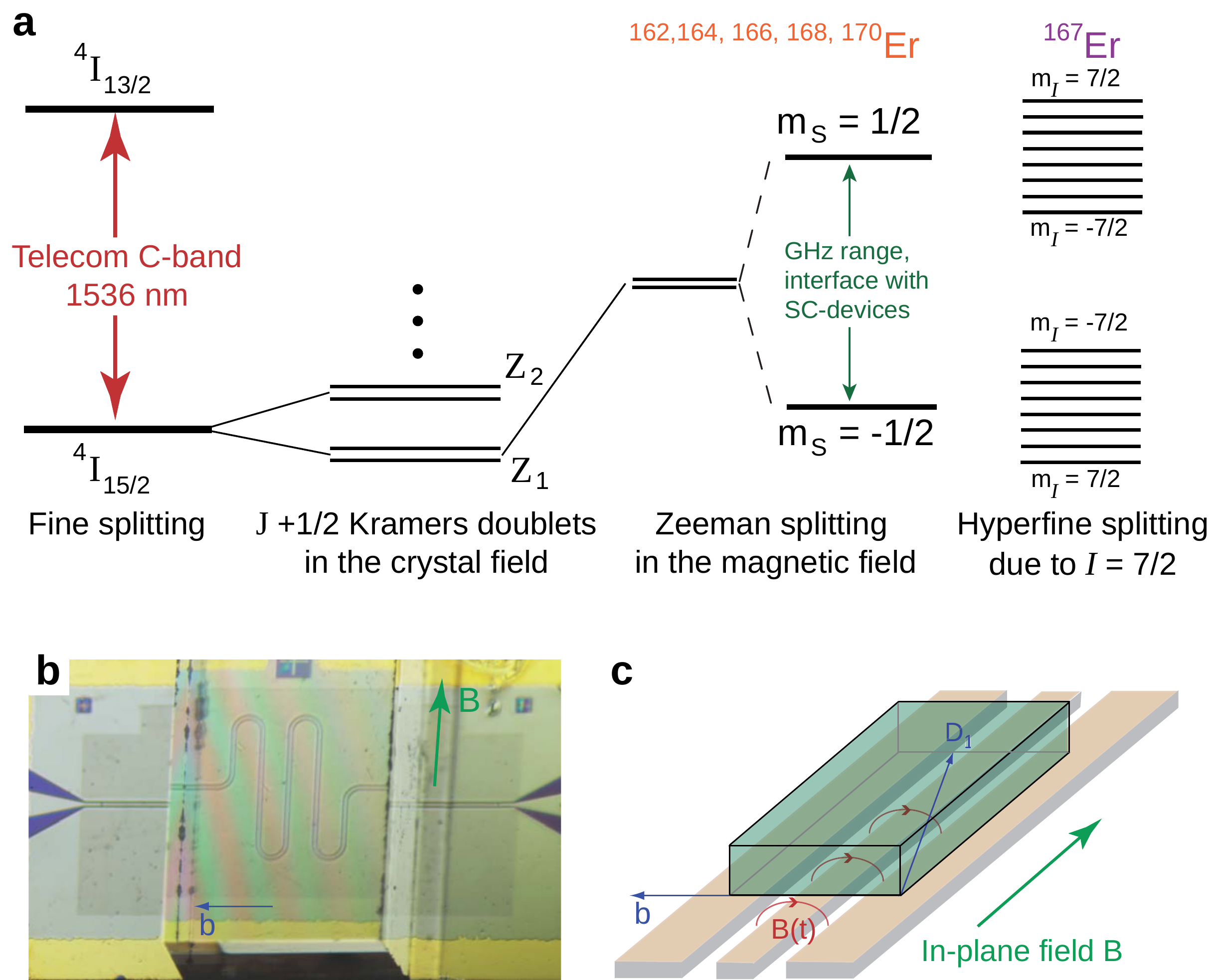}
    	\caption{(Color online)\textbf{(a)}~The structure of the energy levels of Er$^{3+}$ ions. \textbf{(b)}~A picture of the rare-earth ion chip. The Er:Y$_{2}$SiO$_{5}$ crystal of 1 $\times$ 1.5 $\times$ 3 mm$^3$ size is placed on the $\lambda/2$ resonator. The width of the central transmission line of the resonator is 20 $\mu$m; gaps are 9 $\mu$m. The magnetic field is applied along the chip surface. \textbf{(c)}~The sketch of the crystal orientation marked with its optical extinction axis \textbf{b} and \textbf{D}$_1$ with respect to directions of the bias magnetic field B and microwave field B(t). }
	\label{fig:spectrum}
\end{figure}

The energy level diagram of Erbium ions embedded inside a crystal is shown on Fig.\ref{fig:spectrum}(a). The electronic configuration of a free Er$^{3+}$ ion is 4$f^{11}$ with a $^{4}$I term. The spin-orbit coupling splits it into several fine structure levels. An optical transition at telecom wavelength occurs between the ground state $^{2\textrm{s}+1}$L$_{\textrm{J}}=^{4}$I$_{15/2}$ and the first excited state $^{4}$I$_{13/2}$, where S, L and J are spin, orbital and total magnetic momenta of the ion. The weak crystal field splits the ground state into eight (J+1/2) Kramers doublets~\cite{AbragamESR}. At cryogenic temperature, only the lowest doublet Z$_1$ is populated, therefore the system can be described as an effective electronic spin with $S$ = 1/2. However, Erbium has five even isotopes $^{162}$Er,$^{164}$Er,$^{166}$Er,$^{168}$Er and $^{170}$Er, and one odd isotope $^{167}$Er (natural abundance 22.9\%) with a nuclear spin $I$=7/2. Therefore, the electronic states of $^{167}$Er with effective spin projection $m_S$~=~$\pm$1/2 are additionally split into 8 hyperfine levels~\cite{Guillot2006}.

The magnetic properties of Erbium ions are associated with an unquenched total orbital moment J in a crystal field which results in the appearance of a large magnetic moment of nearly 7$\mu_B$, where $\mu_B$ is the Bohr magneton, at particular orientations of the applied magnetic field~\cite{Kurkin1980}. The large spin tuning rate $\sim$~200 GHz/T makes Erbium doped crystals favorable for its integration with SC qubits that can be operated only at relatively low magnetic field. Such an integrated spin-SC device can be very attractive for its applications particularly in quantum repeaters, where one can store quantum information and perform local operations~\cite{Brigel1998}. A quantum state of the optical field can be mapped into spin-waves encoded in Zeeman or hyperfine levels of Erbium ions~\cite{Lukin2001} and transferred later to the quantum state of the microwave field~\cite{Molmer2008,Imamoglu2009}. The strong coupling between SC-qubits and microwave resonator allows fast quantum gate operations on the state of microwave field at nanosecond timescales~\cite{Schoelkopf2009,Martinis2010}. This sequence can also be launched in reverse order, thus establishing a coherent quantum transfer between GHz and optical frequency ranges.

In this experiment, we use a single Y$_{2}$SiO$_{5}$ crystal doped with 0.02\% of Er$^{3+}$ (Er:YSO), supplied by Scientific Materials Inc. The crystal has dimensions of 1~$\times$~1.5 $\times$~3 mm$^3$ and it is glued on top of the silicon chip with a $\lambda$/2-coplanar Niobium SC-resonator, see Fig.\ref{fig:spectrum}(b). The resonance frequency of the rare-earth ion chip is $\omega_0$~=~$2\pi\times$8.9 GHz and its quality factor is $Q\approx$ 32000 at magnetic field of Erbium transitions. The crystal orientation is shown in Fig.\ref{fig:spectrum}(c) with its optical extinction axes \textbf{b} and \textbf{D}$_1$ and is specified by the angles $\theta$ and $\phi$ between these axes and direction of the applied magnetic field~\cite{Sun2008}. The \textbf{b}-axis of the crystal is directed along its 1.5 mm side and is perpendicular to the bias magnetic field B ($\theta$~=~90$^{\circ}$) applied parallel to the chip surface. The angle between the magnetic field and \textbf{D}$_1$ is $\phi$~=~$-$60$^{\circ}$. The particular orientation of our experiment is chosen to maximize the g-factor for the crystallographic site 1~\cite{Guillot2006,Sun2008}, which falls into our field scanning range between 0 and 70 mT and is relevant for our experiment. The gap between the bottom of the crystal and the chip surface is controlled by observing Newton's interference fringes and with proper placement 1-2 fringes are visible, yielding a gap of about 0.5 $\mu$m. The experiment was performed in a $^3$He cryostat, at a base temperature of 280 mK.

In quantum optics, the resonator-spin ensemble interaction is usually described by the Tavis-Cummings Hamiltonian~\cite{Henschel2010}. Provided that number of photons in the resonator is much smaller than number of spins $N$, a spin ensemble behaves as a harmonic oscillator coupled to a cavity. That results in an avoided level crossing when the spins are tuned into resonance with a cavity, which is also in agreement with a phenomenological treatment of the observed effect. The rotating component of the magnetization of the spins produces and oscillatory magnetic field, which perturbs the inductance $L_0$ of the resonator. The new inductance $L'=L_0(1+\chi\xi)$, where $\chi=\chi'-i\chi''$ is the dynamic magnetic susceptibility of the ensemble with $\chi'$ and $\chi''$ being its dispersive and absorptive parts respectively, and $\xi$ is a geometric factor describing the spin distribution across the mode and its coupling to the oscillating field. Thus, the resonance frequency of the rare-earth chip is $\omega'_0 =\omega_0/\sqrt{1+\chi\xi}$. Provided that $\chi\xi \ll$1, we obtain the following equations for the resonator frequency $\omega'_0$ and its decay rate $\kappa'_0$:

\begin{equation}\label{omega'}
\omega'_0 =\omega_0+\frac{v^2~(\omega_0-\omega_s)}{(\omega_0-\omega_s)^2+\Gamma_2^{\ast 2}}~,
\end{equation}
\begin{equation}\label{kappa'}
\kappa'_0 =\kappa_0+\frac{v^2~\Gamma_2^{\ast}}{(\omega_0-\omega_s)^2+\Gamma_2^{\ast 2}}~,
\end{equation}
where $v$ is coupling strength, $\omega_s$ is the Larmor frequency of spin ensemble, and $\Gamma_{2}^{\ast}$ is its total linewidth~\cite{AbragamESR}. The coupling strength $v$ can be expressed via the static susceptibility of the spin ensemble $\chi_0$ and reads as $v=\omega_0\sqrt{\chi_0\xi/2}$. The substitution of the actual value of $\chi_0$ yields the final expression for the collective coupling strength $v=\tilde{\textrm{g}}\mu_B \sqrt{\mu_0 \omega_0 n \xi/4 \hbar}$, where $\tilde{\textrm{g}}$ is an effective \textrm{g}-factor due to the magnetic anisotropy of the crystal and $n$ is the spin concentration. For the concentration of Erbium electronic spins of $n_S\sim$10$^{18}$cm$^{-3}$, the effective \textrm{g}-factor in the plane perpendicular to the magnetic field is $\tilde{\textrm{g}}\sim$~7 and filling factor $\xi\sim$~0.25 we expect the coupling strength $v/2\pi\sim$~60 MHz.

\begin{figure}[htb]
	    	\includegraphics[width=0.8\columnwidth]{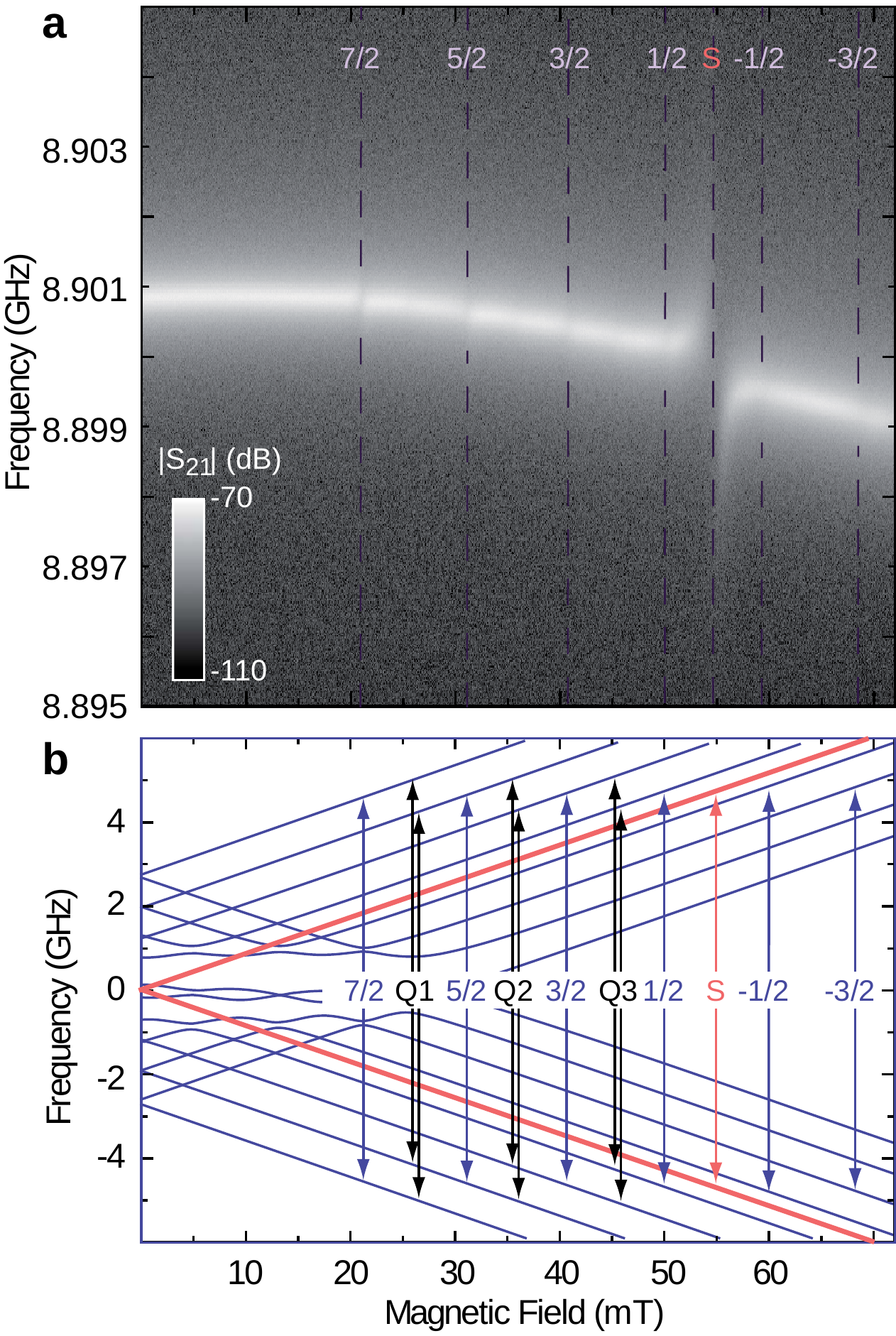}
	    	\caption{(Color online) Transmission spectroscopy of the rare-earth ion chip. (\textbf{a})~The amplitude of the transmitted signal $|S_{21}|$. The energy spectrum of different magnetic transitions is shown with dashed lines. The spin tuning rate $\gamma_S/2\pi$~=~162.9 GHz/T and values for the hyperfine splitting are taken from the fit of the frequency shift(see the text for details). (\textbf{b})~The calculated energy spectrum of Er:YSO in the magnetic field. Allowed and forbidden transition are shown with arrows.}
		\label{fig:results}
\end{figure}

The microwave transmission spectroscopy of the rare-earth ion chip as a function of the bias magnetic field is presented in Fig.\ref{fig:results}. By using a vector network analyzer we measured the $S_{21}$ parameter, which contains both the amplitude and the phase of the signal. The probing power at the input of the resonator is about 1 fW corresponding to an excitation level of $\sim$~100 microwave photons in the SC-resonator. The amplitude of the transmitted signal $|S_{21}|$ as a function of magnetic field and probing frequency is presented in Fig.\ref{fig:results}(a). The spectrum consists of the large avoided level crossing at a bias field of $B$~=~55 mT and associated with magnetic dipole transition between electronic states with $m_S$~=~$\pm$1/2. A regular pattern of six dark interruptions is associated with the odd Erbium isotope and is due to allowed hyperfine transitions between states with equal nuclear spin projection $m_I$~=~7/2, 5/2,..., -1/2, -3/2. The states with $m_I$~=~-5/2 and -7/2 do not enter into our scanning range.

To understand the measured spectrum we numerically diagonalize the spin Hamiltonian for paramagnetic ion in the crystal~\cite{AbragamESR,Guillot2006}:

\begin{equation}\label{ErbiumH}
H = \mu_B \textbf{B} \cdot \textbf{g} \cdot \textbf{S} + \textbf{I}\cdot \textbf{A}\cdot \textbf{S} + \textbf{I}\cdot \textbf{Q}\cdot \textbf{I} - \mu_n \textrm{g}_n \textbf{B}\cdot\textbf{I},
\end{equation}
where $\textbf{g}$ is the g-factor tensor, $\textbf{A}$ is the hyperfine tensor, $\textbf{Q}$ is the nuclear quadrupole tensor, $\mu_n$ is nuclear Bohr magneton and g$_n$ is nuclear g-factor. The first term in the Hamiltonian presents an electronic Zeeman splitting, the second one describes hyperfine interactions, the third one is the quadrupole term and the last one is the Zeeman splitting due to the nuclear spin. The values for the tensors has been taken from a previous ESR study of Er:YSO crystal~\cite{Guillot2006}. The resulting eigenspectrum is presented in Fig.\ref{fig:results}(c). The position of each energy level is drawn as a function of the applied magnetic field, and the experimentally observed magnetic transitions are shown by arrows. For the even Erbium isotopes only the first term in the Hamiltonian~(\ref{ErbiumH}) survives and that results in the strong magnetic transition between electronic spin states $m_S$~=~$\pm$1/2 shown with the red arrow and marked by the letter "S". The odd $^{167}$Er ion has 8 allowed hyperfine transitions when $\Delta m_I$~=~0, which are shown by blue arrows and marked with m$_I$ numbers.

\begin{figure}[htb]
	    	\includegraphics[width=0.9\columnwidth]{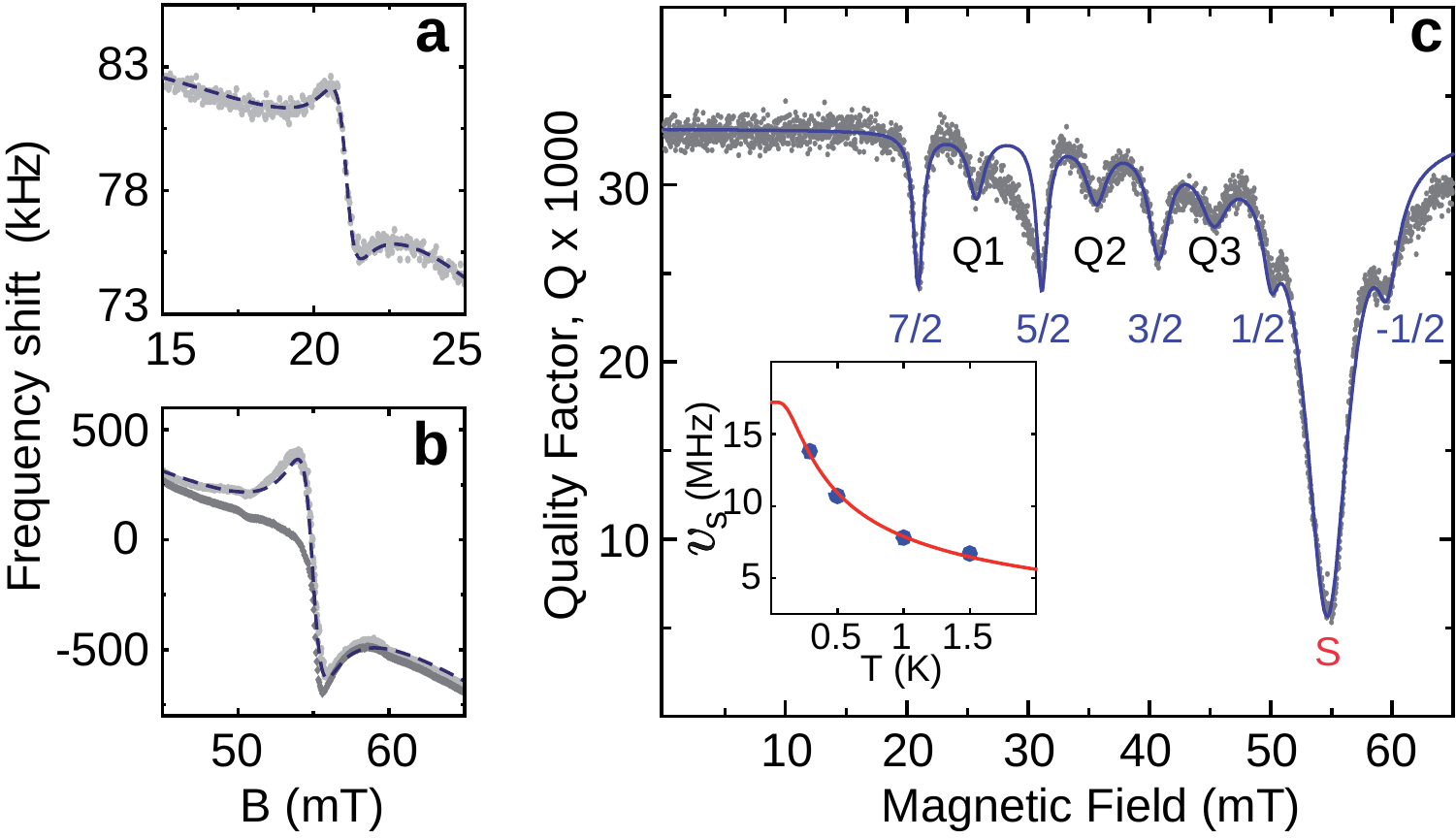}
	    	\caption{(Color online)~Frequency shift of the resonator $\omega'_0/2\pi-$8.9~GHz and its quality factor Q versus the applied magnetic field. \textbf{(a)} For the transition "7/2". Gray dots are experimental data. The dashed line is fit to Eq.(\ref{omega'}).~\textbf{(b)} For the transition "S". Light gray and dark gray dots are experimental data taken at excitation level of 10$^2$ and 10$^5$ photons respectively. The dashed line is a fit of the low excitation data to Eq.(\ref{omega'}).~\textbf{(c)}~Gray dots are experimental data points. The solid line is fit of experimental data. Clearly observed dips corresponds to different magnetic transition. Inset: The temperature dependence of the collective coupling $v_s$ versus temperature. Circles present measured data. The solid line is a fit to the data~(see the text for details).}
		\label{fig:Qfactor}
\end{figure}

To extract the coupling strengths and linewidths for different transitions, each spectral line $|S_{21}|$ is fitted to a Lorentzian at every value of the magnetic field. The data corresponding to the shift of the resonator frequency $\omega'_0 $ for transitions "7/2" and "S" is shown in Fig.\ref{fig:Qfactor}(a),(b). The dispersive behavior of the frequency shift in the vicinity of Erbium spin transitions is well fit with Eg.(\ref{omega'}). For the transition "7/2", the coupling strength extracted from that fit is $v_{7/2}/2\pi$~=~2.1~$\pm$~0.3 MHz, which exceeds the decay rate of the resonator $\kappa/2\pi$~=~0.3 MHz. However the linewidth of spin ensemble exceeds the coupling strength and the same fit yields $\Gamma^{\ast}_{7/2}/2\pi$~=~70~$\pm$~1 MHz. The frequency shift of the resonator due to the coupling to the electronic spin ensemble "S"~( Fig.\ref{fig:Qfactor}(b)) has been studied for two excitation levels of 10$^2$(light gray dots) and 10$^5$(dark gray dots) microwave photons. The dashed line described by Eq.(\ref{omega'}) fits well the low excitation spectrum yielding $v_{S}/2\pi$~=~11.6~$\pm$~0.2 MHz and $\Gamma^{\ast}_{S}/2\pi$~=~145~$\pm$~5 MHz. At high excitation level, the presented data corresponds to a Dysonian line shape associated with spin diffusion in bulk metals~\cite{Dyson1955}. We believe that the observed mixture of the dispersive and absorptive line shapes might also be associated with spin diffusion in and out of the part interacting with the resonator mode.

The measured spectrum in Fig.\ref{fig:results}(a) also reveals other interesting features: between hyperfine transition one can recognize an additional regular structure. To study this, we plot the quality factor of the resonator on the Fig.\ref{fig:Qfactor}(c) as a function of magnetic field. The curve consists of a series of regular absorption dips originating from the magnetic coupling of electronic and hyperfine spin ensembles. It also contains an additional weak pattern appearing in between the hyperfine transitions. We interpret these weak absorption lines as corresponding to the forbidden quadrupole transitions between the hyperfine states satisfying $\Delta m_I =\pm$~1~\cite{AbragamESR,Guillot2006}. These transitions are marked by the letters "Q1", "Q2" and "Q3".

The experimental data on the Fig.\ref{fig:Qfactor} can also be fit with $Q = \omega_0/\kappa'_0$ by using Eq.(\ref{kappa'}), where each magnetic transitions contributes independently to the cavity decay rate $\kappa'_0$. The data points between transitions "Q1" and "5/2", and after transition "-1/2" are not fit well due to the presence of additional magnetic transitions related the other crystallographic site. The fit of the quality factor behavior yields the following values: $v_{7/2}/2\pi$~=~2.5$~\pm$~0.3 MHz, $\Gamma^{\ast}_{7/2}/2\pi$~=~65~$\pm$~3 MHz and $v_{S}/2\pi$~=~13.8~$\pm$~0.1 MHz, $\Gamma^{\ast}_{S}/2\pi$~=~141~$\pm$~3 MHz. The slight differences of fitted parameters by using $\omega'_0$ or $Q$ are probably due to an admixture of two magnetic classes into one anticrossing.

The dimensionless parameter to identify the coupling regime is the cooperativity $C=v^2/\kappa\Gamma^{\ast}$~\cite{Vuletic2011}, which corresponds to the number of coherent oscillations between coupled systems. The cooperativity parameter for the hyperfine transition "7/2" is measured to be $C_{7/2}$~=~0.4. High cooperativity coupling is reached for the electronic transition "S", where $C_S~$=~5.2. In that regime microwave photons inside the SC-cavity coherently interacts with a spin ensemble. However, to observe a normal-mode splitting the condition $2v>(\kappa, \Gamma^{\ast})$ has to be fulfilled.

The coupling strength $v_S$ of the electronic spin ensemble is found to vary as function of its temperature. The transmission spectrum of the rare-earth chip has been taken at temperatures of 0.3, 0.5, 1 and 1.5 Kelvin. The coupling strength $v_S$ is extracted from the change of the \textrm{Q} and plotted as a function of the temperature in the inset of Fig.\ref{fig:results}(b). The temperature dependence is fit to $v_S = v_S(0)~[\tanh (\hbar \omega_0 /2k_B T)]^{1/2}$, see~\cite{AbragamESR}, and plotted with solid line. The coupling strength of the ensemble at zero temperature, $v_S(0)/2\pi$~=~17.3~$\pm$~0.2 MHz, is the only fitted parameter.

We also found that linewidths of different spin ensembles are not the same and grow with magnetic field. We explain this by a small misalignment of the crystal with respect to the magnetic field. When the angle $\theta$ deviates from 90$^{\circ}$, the additional degeneracy due to C$_2$ symmetry of the crystal is lifted and each Erbium transition splits further into two different magnetic subclasses~\cite{Kurkin1980,Guillot2006,Sun2008}. Using the ``Easyspin'' package~\cite{EasySpin} we simulated ESR spectra of the Er:YSO crystal and bounded the maximum misalignment of the crystal to be $\Delta\theta<$0.4$^{\circ}$. Assuming that the observed effect is dominated by the magnetic class splitting, an additional contribution to the spin linewidth is estimated to be about 30 MHz. Such large linewidth can not be completely attributed to the dephasing of spins. We measured the pure dephasing time using a Hahn echo sequence in pulsed ESR spectrometer at the temperature of 7~K, and found that for the "7/2" transition $T_2^{(7/2)}\approx$~540 ns and for the "S" transition $T_2^{(S)}\approx$~200 ns. Moreover, at the temperature of 0.3~K we would expect $T_2$ to be longer by at least an order of magnitude~\cite{Bertaina2007,Awschalom2008}. Therefore, the additional contribution to the spin linewidths may be associated with field inhomogeneity or surface magnetism of the SC-resonator~\cite{Sendelbach2009}.


In conclusion, we have presented measurements on an Er$^{3+}$ spin ensemble in Y$_2$SiO$_5$ crystal magnetically coupled to a SC coplanar resonator. This hybrid system is a promising candidate for an interface between SC quantum circuits and optical quantum networks. The measured coupling strengths of different spin ensembles encoded in Er:YSO crystal exceeds the decay rate of the cavity. The demonstrated on-chip ESR spectroscopy allows us to resolve all spin state transitions of the Er$^{3+}$ ions. The presented experiment realizes a first step towards implementation of a rare earth based quantum memory with telecom wavelength conversion that could constitute an essential building block of future quantum networks.

We thank M. Azarkh and M. Drescher for the ESR study of Er:YSO crystals, A. Abdumalikov and O. Astafiev (NEC) for the fabrication of the SC-resonator and C. M\"{u}ller for his useful comments and critical reading of the manuscript. This work was supported by the CFN of DFG, the EU projects MIDAS, SCOPE, SOLID and the BMBF programm "Quantum communications". IP acknowledges support of Alexander von Humboldt Foundation. PB acknowledges the financial support of RiSC grant of KIT and Baden-W\"{u}rttemberg.

\bibliographystyle{apsrev}

\bibliography{Erbium_QED}

\end{document}